\documentclass[11pt,twocolumn]{article}

\usepackage{graphicx}
\usepackage{algorithm,algorithmic}
\usepackage{amsmath}
\usepackage{multirow}
\usepackage{float}
\usepackage{url}
\usepackage[none]{hyphenat}
\usepackage{authblk}

\usepackage{txfonts,balance}

\usepackage{afterpage}

\begin{document}
\newcounter{cntr1}
\newcounter{cntr2}
\emergencystretch 3em

\title{An open-source simulation package for power electronics education}

\author[1]{Mahesh~B.~Patil}
\author[2]{V.V.S.~Pavan~Kumar~Hari}
\author[1]{Ruchita~D.~Korgaonkar}
\author[1]{Kumar~Appaiah}
\affil[1]{Department of Electrical Engineering, Indian Institute of Technology Bombay}
\affil[2]{Department of Energy Science and Engineering, Indian Institute of Technology Bombay}

\maketitle

\begin{abstract}
Extension of the open-source simulation package GSEIM\,\cite{patil2021gseim}
for power electronics applications is presented. Recent
developments in GSEIM, including those oriented specifically
towards power electronic circuits, are described. Some examples
of electrical element templates, which form a part of the GSEIM
library, are discussed. Representative simulation examples in
power electronics are presented to bring out important features
of the simulator.
Advantages of GSEIM for educational purposes are
discussed. Finally, plans regarding future developments in
GSEIM are presented.
\end{abstract}

\section{Introduction}
Simulation can be a very effective tool for improving students'
understanding of fundamental concepts since it allows verification
of the concepts as well as quick exploration of several
``what-if" scenarios. In the context of power electronics, for example,
simulation allows the student to view the effect of changing a duty
ratio or an inductance value on the voltage and current waveforms
in the circuit under discussion, thus reinforcing the concepts being
taught in class. Several commercial simulation tools are currently
being used for teaching power electronics, including
PSIM\,\cite{psim},
PSCAD\,\cite{pscad},
Matlab Simulink/Simscape\,\cite{simulink}, and
PLECS\,\cite{plecs}.
While academic versions of these packages at a lower cost or
free student versions (with limitations) are generally available,
open-source options are certainly advantageous, especially for
engineering colleges in developing countries.

There are currently few open-source options for power electronics.
Of those, Openmodelica\,\cite{openmodelica} is based on the hardware description
language Modelica, while GeckoCIRCUIT\,\cite{musing2014successful}
is a java-based platform. Open-source tools are currently not being used
for power electronics education on a large scale probably because of attractive
features such as ease of use and customer support associated with
commercial packages.

An open-source simulator GSEIM was recently reported\,\cite{patil2021gseim}.
In the first version, GSEIM was aimed at simulation of power electronic
systems which can be represented by a flow-graph, e.g., V/f control of
an induction motor. Subsequently, GSEIM has been extended, both in terms
of GUI features and numerical engine, to enable simulation of a number
of power electronic circuits covered in typical undergraduate and
postgraduate courses. It is the purpose of this paper to report the
current status of the GSEIM package and point out its potential as an
open-source tool for power electronics education.

The paper is organised as follows.
In Sec.~\ref{sec_recent}, recent developments in GSEIM are reported.
The most important development, viz., addition of electrical elements
in the form of templates, is described in Sec.~\ref{sec_ebe}, with
the help of examples. In Sec.~\ref{sec_examples}, simulation examples
are presented to bring out the scope and capabilities of the program.
Advantages of GSEIM as an open-source package have been pointed
out in Sec.~\ref{sec_open}.
Finally, in Sec.~\ref{sec_conclusions}, conclusions of this work
are summarised, and future developments envisaged in GSEIM are listed.

\section{Recent developments in GSEIM}
\label{sec_recent}
The currently available GSEIM program\,\cite{gseimgithub}
allows the user to enter the schematic diagram of the system
of interest using a graphical user interface (GUI), specify
component values, run simulation, and plot results interactively.
In addition, it allows the user to create new elements (blocks)
either in terms of equations or as hierarchical blocks made up
of elements already available in the library. Applications are
limited to power electronic systems which can be represented as
a flow-graph, with each element having input and output nodes.
The primary objective of the new GSEIM version presented in this
paper is to allow simulation of electrical circuits. For convenience,
we will call the new GSEIM version GSEIM-Electrical (GSEIM-E) and the
original GSEIM program described in \cite{patil2021gseim} as
GSEIM-Flowchart (GSEIM-F). In the following, we summarise the salient
features of GSEIM-E.
\begin{list}{\it\Alph{cntr1}.}{\usecounter{cntr1}}
 \item
  Numerical engine: The numerical engine (C++) of the GSEIM-F program
  was extended to handle electrical elements. The modified nodal
  analysis (MNA) approach, along with the Newton-Raphson method
  for nonlinear circuits, was implemented. When the system being
  simulated has electrical elements, only implicit methods~-- backward
  Euler or trapezoidal method with constant or variable
  time steps~-- are allowed for numerical integration. The details
  of these techniques can be found in \cite{sequelmbp} and references
  therein.
 \item
  Steady-state waveform (SSW) analysis: In several converter
  applications, the steady-state waveform is of interest.
  In priniciple, transient simulation performed for a sufficiently large number
  of cycles can yield the steady-state solution. However,
  this process can take too long if the circuit time constants are
  large. The Newton-Raphson time-domain steady-state waveform (NRTDSSW)
  method described in \cite{patil2002ssw} is implemented in GSEIM-E
  for directly obtaining the steady-state solution. An example would
  be presented in Sec.~\ref{sec_examples}.
 \item
  Rectilinear wiring and electrical nodes: In GSEIM-F, the GUI was
  built by making suitable changes in the GNURadio\,\cite{gnr}
  GUI, and like its predecessor, the GSEIM-F GUI allowed only
  curved wires (using splines). For electrical circuits, rectilinear
  wires were incorporated in GSEIM-E, and electrical nodes (ports)
  were added.
 \item
  Element symbols: In GSEIM-F, elements (blocks) were displayed
  using rectangles, with the type of the element appearing inside
  the rectangle. In GSEIM-E, circuit symbols, such as resistor and
  capacitor, are also incorporated. A symbol is rendered in the GUI
  using a python file associated with that symbol. The user can add
  a new symbol by simply adding a python file with a suitable name,
  without making any changes in the GUI code.

  Fig.~\ref{fig_cap}
  shows the python file associated with the capacitor symbol. The
  code between
  {\tt{\#begin\_cord}} and
  {\tt{\#end\_cord}}
  prepares the points involved in the symbol, and the code between
  {\tt{\#begin\_draw}} and
  {\tt{\#end\_draw}}
  does the rendering.
\begin{figure}[!ht]
\centering
\scalebox{0.8}{\includegraphics{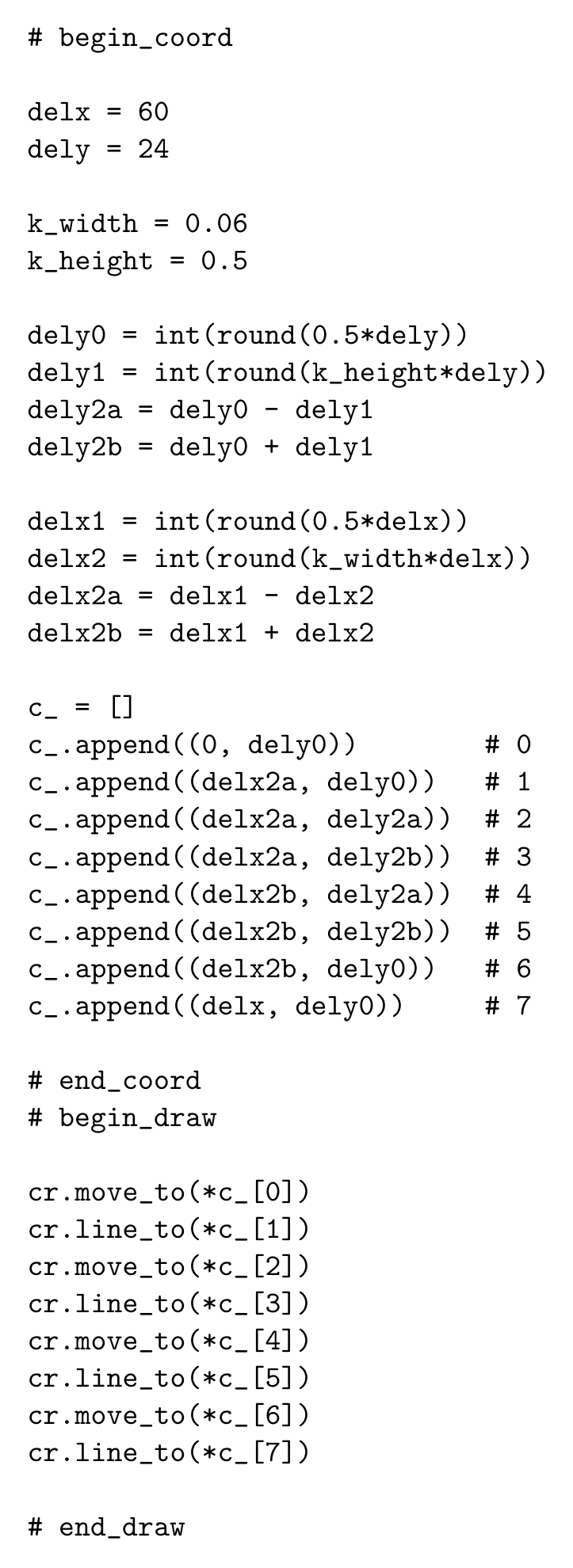}}
\vspace*{-0.2cm}
\caption{Python file associated with the capacitor symbol.}
\label{fig_cap}
\end{figure}
 \item
  Plotting: The following post-processing features have been added to
  the plotting GUI:
  \begin{list}{(\roman{cntr2})}{\usecounter{cntr2}}
   \item
    Average and rms values
   \item
    Fourier spectrum and total harmonic distortion (THD)
  \end{list}
\end{list}
In some aspects, GSEIM is similar to SEQUEL\,\cite{sequelmbp}. However,
the organisation of GSEIM is significantly different. In particular,
the SEQUEL library involves both basic and compound elements,
whereas the GSEIM library involves only basic elements, the compound
elements being treated through the hierarchical block facility provided
by the GSEIM GUI. The other important difference is that GSEIM is oriented
mainly toward power electronics while SEQUEL is more general.

\section{Electrical element templates}
\label{sec_ebe}
As discussed in \cite{patil2021gseim}, the equations governing
the behaviour of an element is incorporated in GSEIM in the form
of ``templates." Some of the flow-graph type element templates
have been discussed in \cite{patil2021gseim}. Here, we look at
a few electrical basic element ({\tt{ebe}}) templates.
\begin{list}{\it\Alph{cntr1}.}{\usecounter{cntr1}}
 \item
  Resistor: Fig.~\ref{fig_r_ebe} shows the resistor template.
  The terminal currents are given by
  $i_p \,$=$\, \displaystyle\frac{v_p-v_n}{R}$,
  $i_n \,$=$\, -\,\displaystyle\frac{v_p-v_n}{R}$.
  The derivatives of these functions
  $\displaystyle\frac{\partial i_p}{\partial v_p}$,
  $\displaystyle\frac{\partial i_p}{\partial v_n}$, etc.
  are constants, and that is indicated by the {\tt{Jacobian}}
  statement.
  The {\tt{nodes}} and the {\tt{rparms}} statements specify
  the nodes and real parameters of the element, respectively.
  The {\tt{outparms}} statement specifies the quantities made
  available by this element to the user for plotting.
\begin{figure}[!ht]
\centering
\scalebox{0.8}{\includegraphics{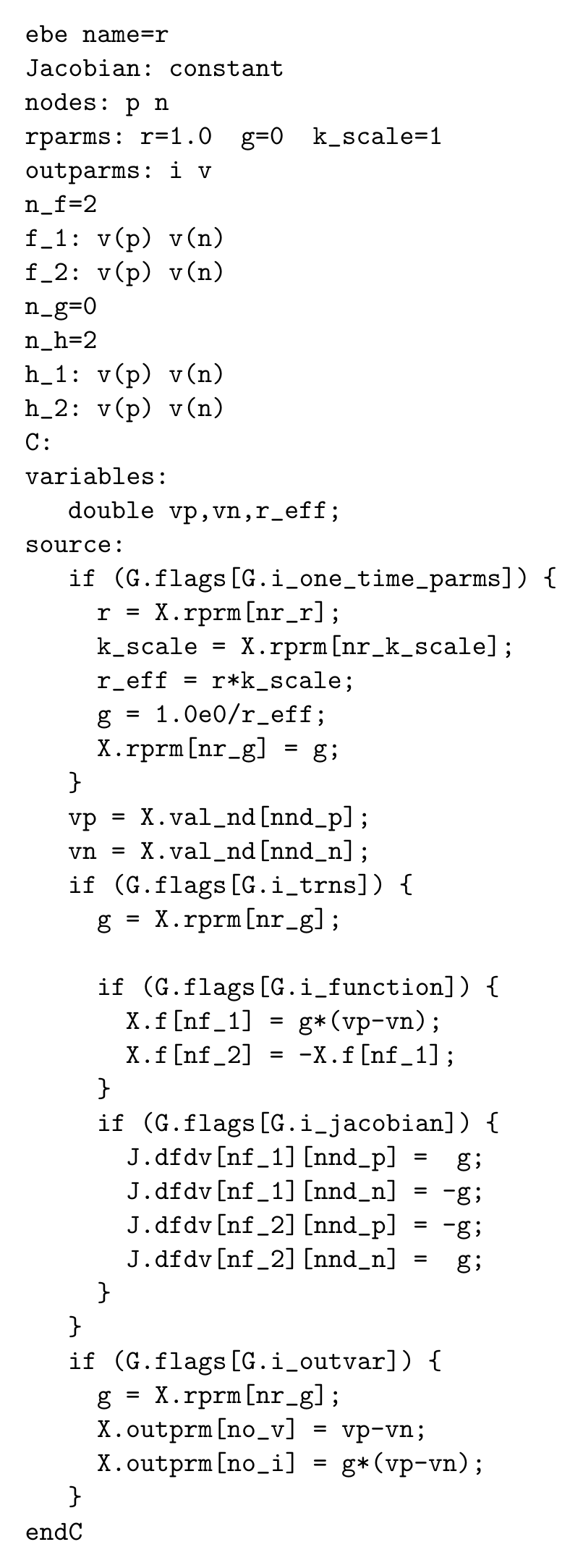}}
\vspace*{-0.2cm}
\caption{Resistor template (partial).}
\label{fig_r_ebe}
\end{figure}

  The main program expects three types of functions to be
  supplied by an electrical element template.
  \begin{list}{(\alph{cntr2})}{\usecounter{cntr2}}
   \item
    Functions
    $f_1$, $f_2$, $\cdots$
    are related to terminal currents in transient simulation.
    If the element has $N$ nodes, the first $N$ of these equations
    give the node currents, while the remaining equations (if any)
    are auxiliary equations.
   \item
    Functions
    $g_1$, $g_2$, $\cdots$
    are related to state variables, as we will see with
    respect to the capacitor template.
   \item
    Functions
    $h_1$, $h_2$, $\cdots$
    are related to ``start-up" simulation, which involves solving
    the circuit equations while holding state variables such as
    capacitor voltages and inductor currents constant, at some
    specified values\,\cite{sequelmbp}. For a resistor, there are
    no state variables, and therefore the $f$ and $h$ equations are
    identical.
  \end{list}
  The statement {\tt{n\_f=2}} conveys that there are two $f$ functions
  for this element. The statements starting with
  {\tt{f\_1:}} and
  {\tt{f\_2:}}
  indicate which variables these functions depend on.
  The main program passes two objects to the template:
  (a)~{\tt{X}} which carries information about the specific element
  being called, and
  (b)~{\tt{G}} which carries global information such as the current
  time point.
  By checking the flags of {\tt{G}}, the template computes appropriate
  quantities, and passes them to the main program by assigning
  suitable variables of {\tt{X}}. Some flags of {\tt{G}} are listed below.
  \begin{list}{(\alph{cntr2})}{\usecounter{cntr2}}
   \item
    {\tt{i\_one\_time\_parms}}: compute ``one-time" parameters which are not
    required to be computed in every time step.
   \item
    {\tt{i\_trns}}: transient simulation
   \item
    {\tt{i\_function}}: compute function values
   \item
    {\tt{i\_jacobian}}: compute jacobian values
   \item
    {\tt{i\_outvar}}: compute output parameters
  \end{list}
 \item
  Capacitor: A capacitor involves a time derivative and therefore
  calls for a very different treatment as compared to a resistor.
  The terminal currents can be written as
  $i_p \,$=$\, \displaystyle\frac{dQ_p}{dt}$,
  $i_n \,$=$\, \displaystyle\frac{dQ_m}{dt}$,
  where $Q_p \,$=$\, C\,(v_p-v_n)$ and $Q_m \,$=$\, -Q_p$ are state
  variables. The functions
  $f_1$, $f_2$, $g_1$, $g_2$
  in the capacitor template shown in
  Fig.~\ref{fig_c_ebe} are used to implement these equations.
\begin{figure}[!ht]
\centering
\scalebox{0.8}{\includegraphics{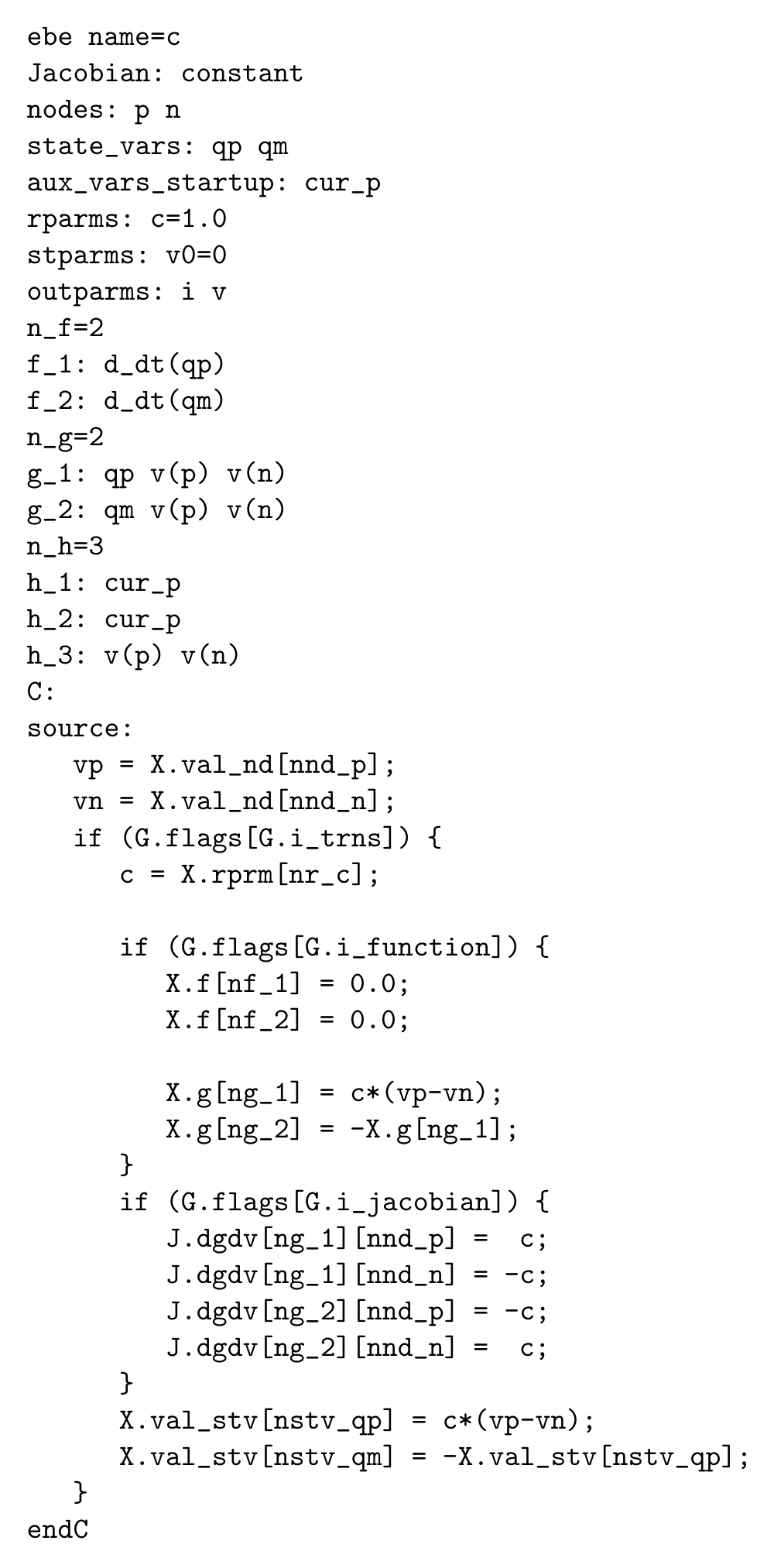}}
\vspace*{-0.2cm}
\caption{Capacitor template (partial).}
\label{fig_c_ebe}
\end{figure}
  In start-up simulation, the capacitor behaves like a dc voltage
  source, satisfying the equations,
  $i_p \,$=$\, i_1$,
  $i_n \,$=$\, -i_1$, and
  $v_p-v_n \,$=$\, V_0$, where the current $i_1$ is an auxiliary
  variable, and $V_0$ is a start-up parameter. Implementation of
  these equations is shown in the start-up part of the capacitor
  template (see Fig.~\ref{fig_c_ebe_1}) where the variable
  {\tt{cur\_p}} is used to denote $i_1$.
\begin{figure}[!ht]
\centering
\scalebox{0.8}{\includegraphics{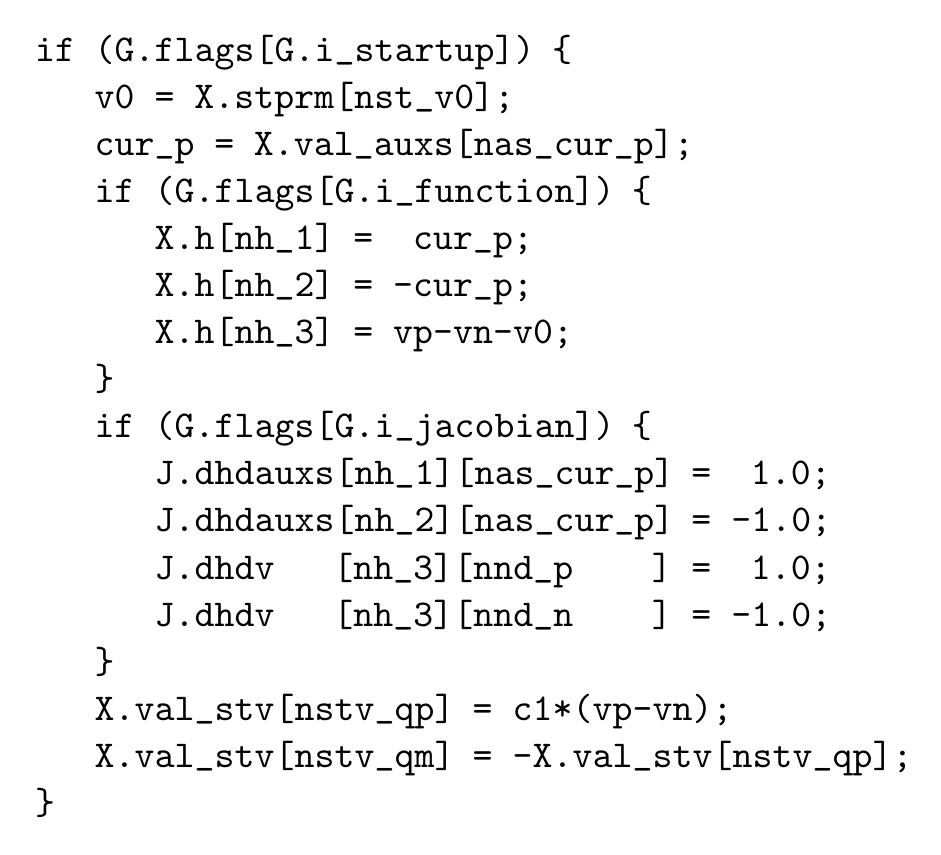}}
\vspace*{-0.2cm}
\caption{Start-up section of the capacitor template.}
\label{fig_c_ebe_1}
\end{figure}
\end{list}

\section{Simulation examples}
\label{sec_examples}
We now present a few simulation examples to demonstrate the
capabilities of GSEIM-E.
As explained in \cite{patil2021gseim}, simulation of a circuit
with GSEIM-F involves drawing the circuit, assigning component
values, setting output variables for plotting, and preparing
an appropriate ``solve block" to specify parameters related to
a specific simulation. This procedure remains the same for
GSEIM-E except for minor changes to handle electrical elements.
The details would be explained in the on-line GSEIM-E documentation,
currently under preparation.

The circuit schematics shown in this
section are taken directly from the GSEIM-E GUI, by exporting
them to {\tt{pdf}} files. Apart from the {\tt{pdf}} format,
the GSEIM-E GUI, like its predecessor GNURadio,
also allows circuit schematics to be exported in
{\tt{svg}} and {\tt{png}} formats. This feature is useful in
preparing presentations or reports.

\begin{list}{\it\Alph{cntr1}.}{\usecounter{cntr1}}
 \item
  $V/f$ control of an induction motor: This example has been described
  in \cite{patil2021gseim}. Here, we show only the schematic diagram
  as it appears in the GSEIM-E GUI in order to demonstrate some of the
  new features of the GUI, viz, rectilinear wiring and the use of
  element symbols.
\begin{figure*}[!ht]
\centering
\hspace*{-0.1cm}{\includegraphics[width=1.0\textwidth]{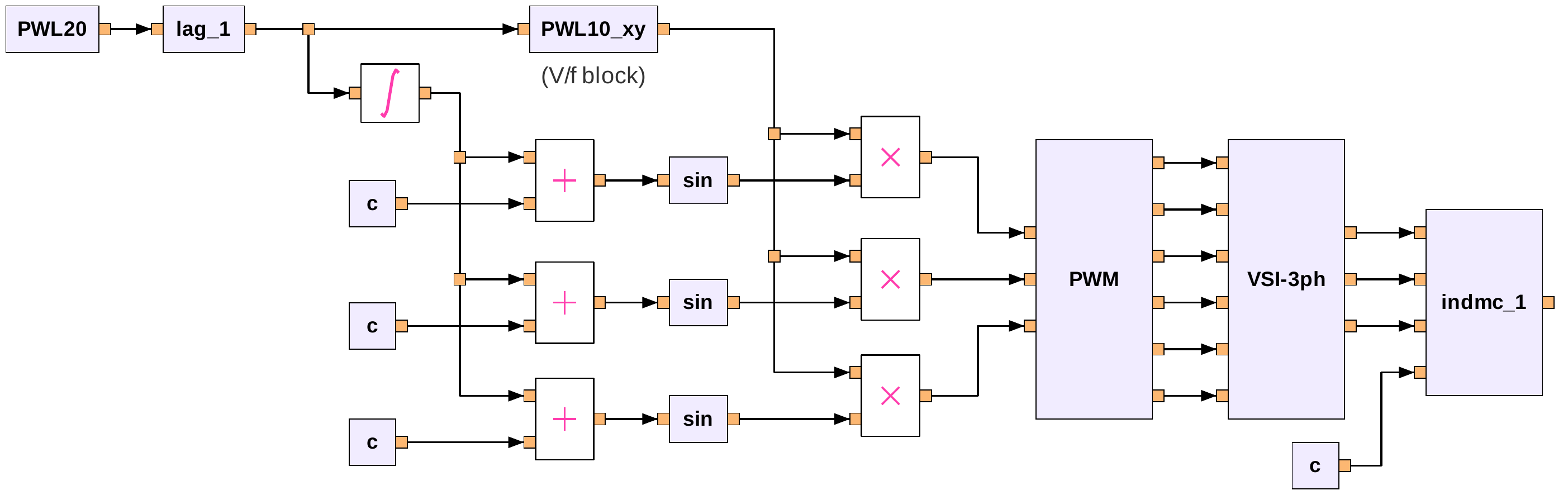}}
\vspace*{-0.2cm}
\caption{Schematic diagram for $V/f$ control of an induction motor.}
\label{fig_vbyf_ckt}
\end{figure*}
 \item
  Buck converter: The buck converter circuit, shown in
  Fig.~\ref{fig_buck_ckt} was simulated for different values
  of duty ratio $D$ and inductance $L$. In each case,
  $i_L \,$=$\, 0$\,A and
  $v_C \,$=$\, 0$\,V
  is taken as the starting point. The output voltage
  $V_o(t)$ is plotted in Fig.~\ref{fig_buck_vout} for three
  cases.
\begin{figure}[!ht]
\centering
\scalebox{0.6}{\includegraphics{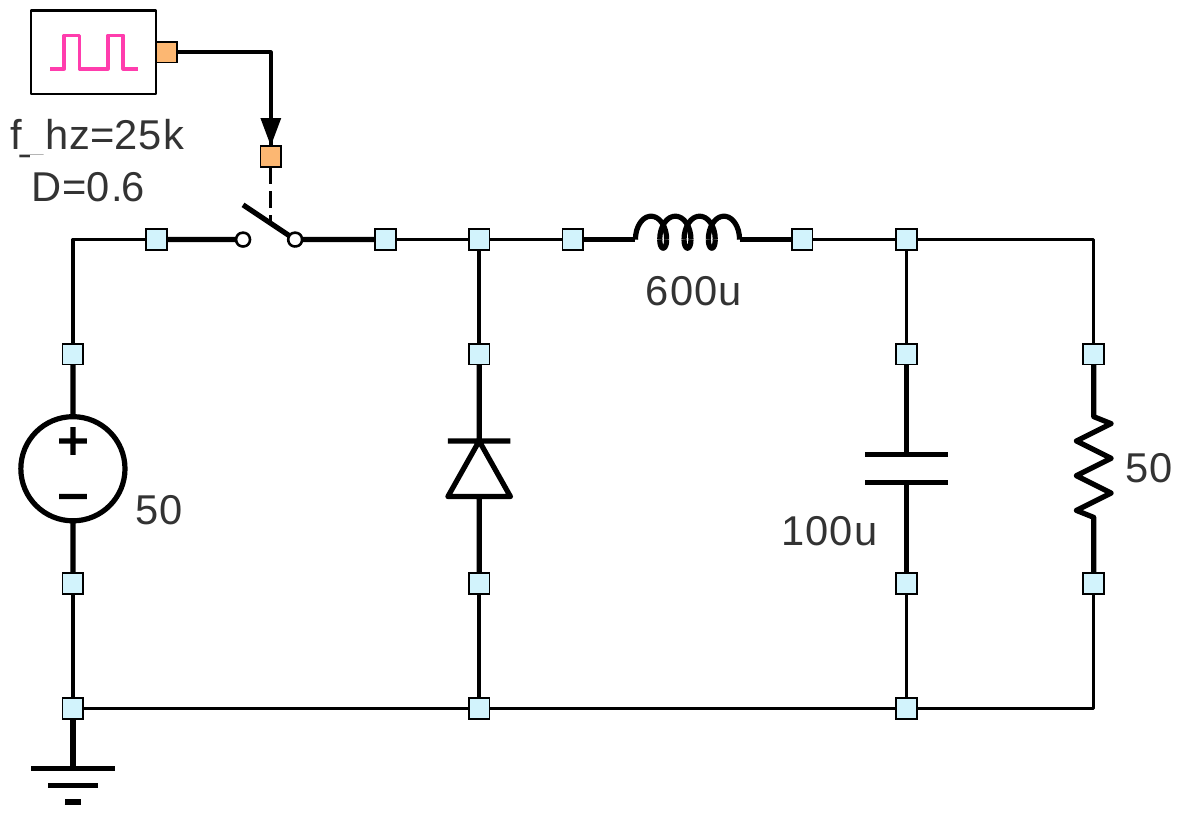}}
\vspace*{-0.2cm}
\caption{Schematic diagram of buck converter.}
\label{fig_buck_ckt}
\end{figure}
  As seen from the figure, the output voltage takes some time
  to settle to its steady-state value.
  Typically, when teaching a
  power electronics course, the steady-state situation is of
  interest, and not the trajectory of the circuit to the steady state.
  Following the transient simulation approach for this circuit~-- and
  also several other converter circuits~-- is therefore wasteful.
  From Fig.~\ref{fig_buck_vout}, we see that, for the component values
  specified, the circuit takes about 10\,msec or 250 cycles to reach
  the steady state, whereas only the last one cycle is of interest.
\begin{figure}[!ht]
\centering
\scalebox{0.68}{\includegraphics{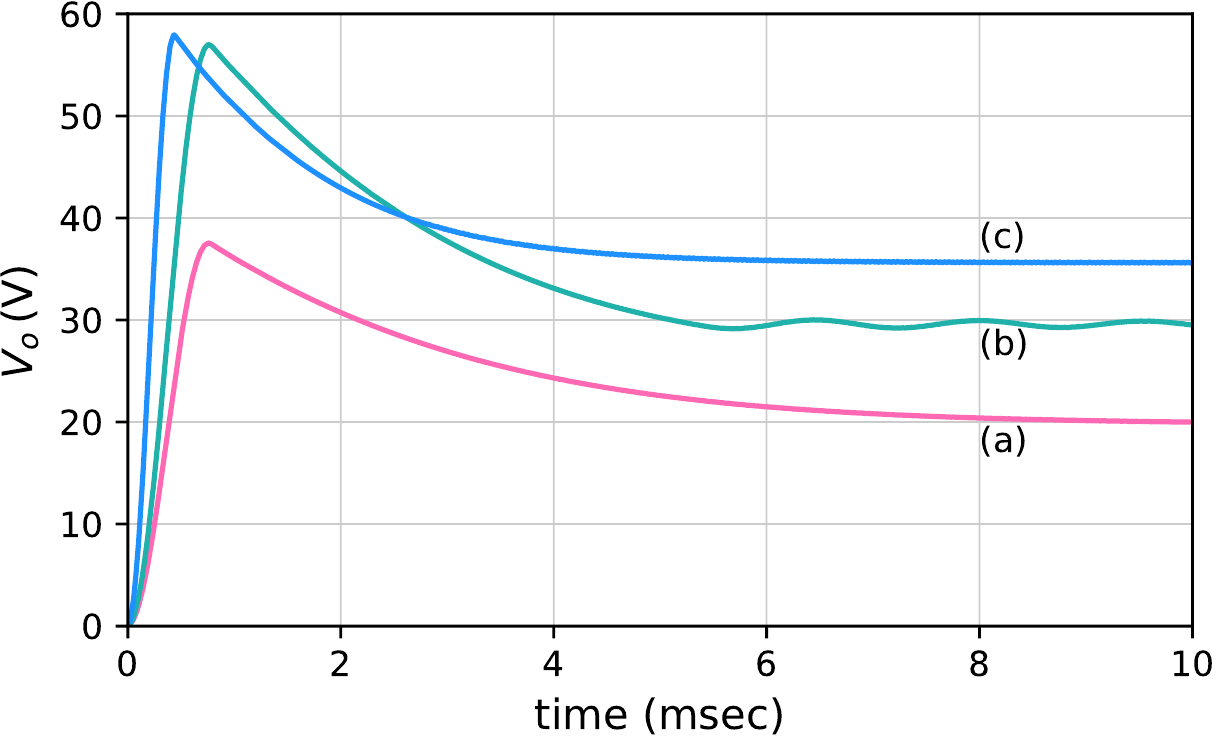}}
\vspace*{-0.2cm}
\caption{Output voltage versus time for the buck converter of Fig.~\ref{fig_buck_ckt}.
(a)\,$D \,$=$\, 0.4$, $L \,$=$\, 600\,\mu$H,
(b)\,$D \,$=$\, 0.6$, $L \,$=$\, 600\,\mu$H,
(c)\,$D \,$=$\, 0.6$, $L \,$=$\, 200\,\mu$H.}
\label{fig_buck_vout}
\end{figure}
  Furthermore, the time taken to reach the steady state depends on the
  parameter values and is generally not know {\it{a priori}}.
  In the classroom, if a teacher wants to demonstrate, for example,
  continuous and discontinuous conduction by changing $L$ or $C$ or $D$,
  transient simulation is clearly not a good option, and a method to
  directly obtain the steady-state solution is desirable.

  GSEIM-E incorporates the Newton-Raphson time-domain steady-state waveform
  (NRTDSSW) approach described in \cite{patil2002ssw} for steady-state
  waveform (SSW) computation. GSEIM-E SSW results for the inductor current
  are shown in Fig.~\ref{fig_buck_il} for the same parameter sets as in
  Fig.~\ref{fig_buck_vout}.
\begin{figure}[!ht]
\centering
\scalebox{0.68}{\includegraphics{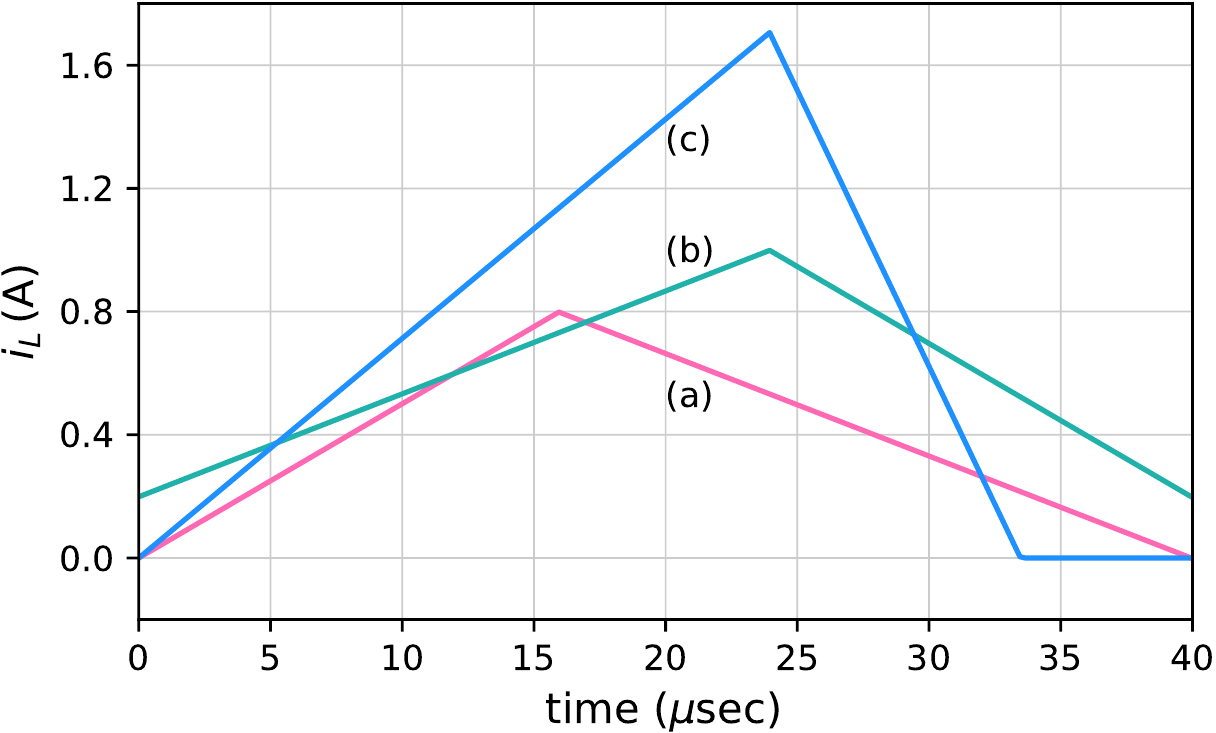}}
\vspace*{-0.2cm}
\caption{Steady-state inductor current versus time for the buck converter of Fig.~\ref{fig_buck_ckt}.
(a)\,$D \,$=$\, 0.4$, $L \,$=$\, 600\,\mu$H,
(b)\,$D \,$=$\, 0.6$, $L \,$=$\, 600\,\mu$H,
(c)\,$D \,$=$\, 0.6$, $L \,$=$\, 200\,\mu$H.}
\label{fig_buck_il}
\end{figure}

  To summarise, the SSW approach offers two major advantages over transient
  simulation: (a)\,it is much faster, (b)\,it does not require the user
  to guess the number of cycles required to reach the steady state.
  We expect the SSW feature of GSEIM-E to become one of its most useful
  features for power electronics education. To the authors' best knowledge,
  SEQUEL \cite{sequelmbp} and PLECS \cite{plecs} are the only other
  simulation packages with direct SSW computation capability in the
  context of power electronic circuits.
 \item
  Neutral point clamped inverter:
  Fig.~\ref{fig_npc} shows the schematic diagram of a neutral clamped
  inverter. The clock generation blocks and the switch-diode blocks
  are implemented using subcircuits (hierarchical blocks).
  For this circuit, the Fourier spectrum of the load current
  is of interest. Fig.~\ref{fig_npc_plot} shows the spectrum for the
  load current, as obtained with GSEIM's plotting GUI.
\begin{figure*}[!ht]
\centering
\scalebox{0.5}{\includegraphics{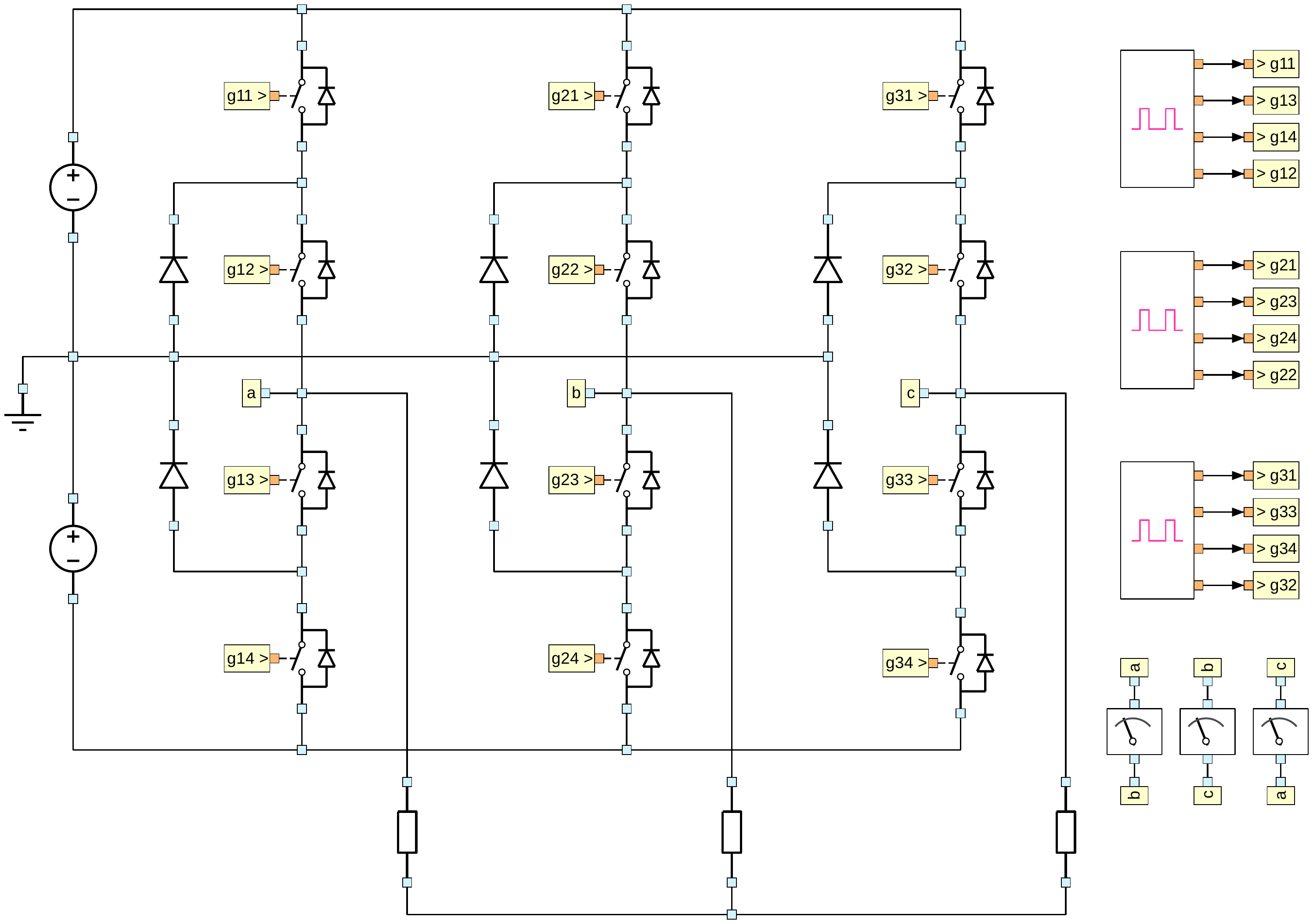}}
\vspace*{-0.2cm}
\caption{Schematic diagram of neutral point clamped inverter.}
\label{fig_npc}
\end{figure*}

\begin{figure}[!ht]
\centering
\scalebox{0.55}{\includegraphics{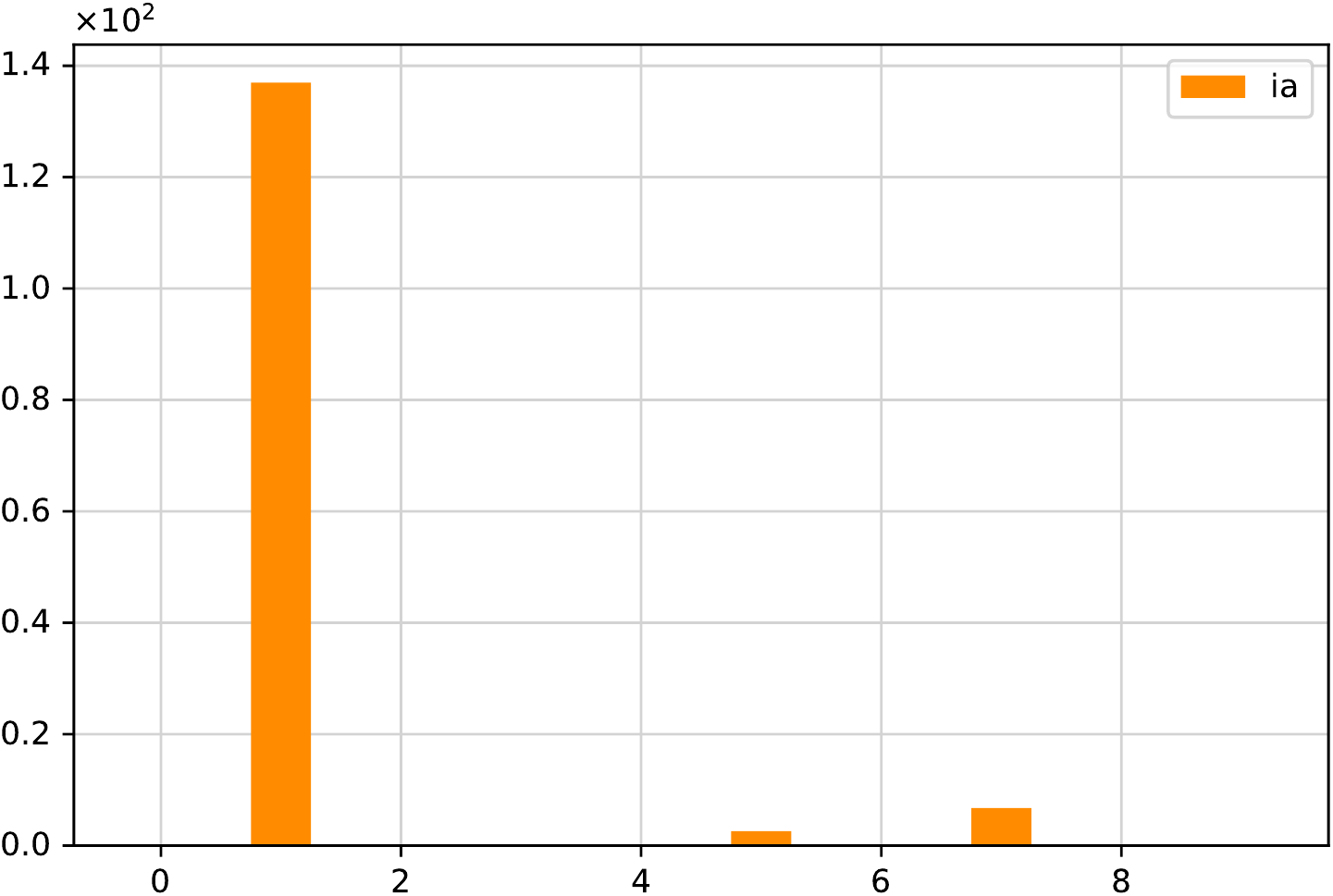}}
\vspace*{-0.2cm}
\caption{Fourier spectrum for neutral point clamped inverter of Fig.~\ref{fig_npc}.}
\label{fig_npc_plot}
\end{figure}
\end{list}
The above examples, together with the machine control examples presented in
\cite{patil2021gseim}, represent the current focus and scope of GSEIM.
We have been able to perform simulation speed comparison for a
set of problems involving electrical machines. We found GSEIM to be 2 to 5
times faster than Simulink in this study\,\cite{akhil}. A more detailed
comparison with Simulink/Simscape and other commonly used commercial
packages is planned; the results will be presented elsewhere.

\section{GSEIM as an open-source package}
\label{sec_open}
The examples presented in Sec.~\ref{sec_examples} bring out the potential
of GSEIM in teaching power electronics courses. Furthermore, the open-source
nature of GSEIM is advantageous in several ways:
\begin{list}{\it\Alph{cntr1}.}{\usecounter{cntr1}}
 \item
  Vendors of commercial packages are constrained from revealing several
  implementation details. As a result, their documentation is mostly about
  ``know-how" rather than ``know-why". Creators of open-source packages are not limited
  by the need for intellectual property protection and can therefore afford to make their
  documentation richer and academically far more rewarding for the users.

  As an example, consider the thyristor block from Simscape\,\cite{thyristor}
  as shown in Fig.~\ref{fig_thyr}. The purpose served by the inductor
  $L_{\mathrm{on}}$ is not explained. Apart from that, consider the following statements
  in the documentation for this block:
  \begin{list}{(\alph{cntr2})}{\usecounter{cntr2}}
   \item
    ``The Inductance {\tt{Lon}} parameter is normally set to 0
    except when the Resistance {\tt{Ron}} parameter is set to 0."
   \item
    ``The Thyristor block cannot be connected in series with an inductor,
    a current source, or an open circuit, unless its snubber circuit is in use."
  \end{list}
  From the user's perspective, these statements appear esoteric and create the
  (wrong) impression that circuit simulation is very complex. On the other hand,
  if the reasons behind these limitations were explained, it would have led to
  a far better understanding of the simulation process.
\begin{figure}[!ht]
\centering
\scalebox{0.8}{\includegraphics{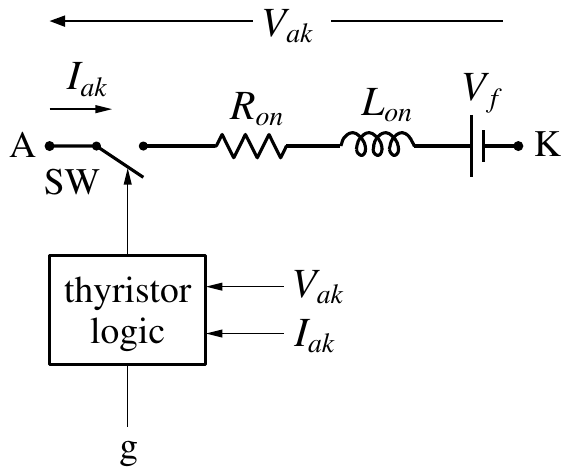}}
\caption{Thyristor block from Simscape\,\cite{thyristor}.}
\label{fig_thyr}
\end{figure}
 \item
  Contributions from users to library elements and simulation examples
  can be easily incorporated in an open-source package like GSEIM. Indeed,
  the basic philosophy behind open-source packages is user involvement in
  not only using the package but also its continuous evolution. In the
  development of GSEIM, special care has been taken in order to allow
  users' contribution in terms of new basic elements, element symbols,
  hierarchical blocks (subcircuits), simulation examples as well as documentation,
  with the hope that the package will grow into a valuable resource for
  power electronics education.
 \item
  Commercial packages often tend to hide, apart from implementation details,
  even data files created by the package, thus forcing the user to use a
  commercial tool~-- generally a part of the same package~-- for viewing the
  results. Open-source packages on the other hand are generally designed
  keeping in mind free exchange of the output files generated by the package
  in {\tt{ASCII}} or {\tt{csv}} format, for example. GSEIM creates output
  files in {\tt{ASCII}} format, and they can be viewed not only with the
  plotting GUI provided with GSEIM, but also with any other plotting program
  including open-source programs like {\tt{gnuplot}} and {\tt{matplotlib}}.
 \item
  Students in several engineering colleges, particularly in developing
  countries, cannot afford licenses for commercial packages. As a
  consequence, teachers are unable to assign home-work exercises involving
  simulation, and students are deprived of the precious learning experience
  offered by simulation. Open-source packages completely remove this
  constraint since no licenses are involved.
 \item
  Open-source packages can take advantage of other open-source tools such
  as compilers, libraries, and plotting programs. This can lead to
  improved capabilities, performance, and implementation.
 \item
  Open-source packages can be combined with other open-source packages
  to create new capabilities at no cost to the user.
  For example, GSEIM can be easily called by an optimisation package
  for circuit design.

  On the other hand, if two commercial packages are combined, the user
  has to pay for each of them, e.g., see \cite{plecssimulink}.
\end{list}

\section{Conclusions and future plans}
\label{sec_conclusions}
GSEIM, an open-source simulation package for power electronics education,
has been presented in this paper. The organisation and features of GSEIM
have been described. Incorporation of new elements in the GSEIM library
has been discussed with the help of specific examples. A few simulation
examples have been considered to illustrate the potential of GSEIM in
power electronics education. Future plans for GSEIM include the following.
\begin{list}{(\alph{cntr2})}{\usecounter{cntr2}}
 \item
  manual preparation and uploading of the revised GSEIM version on
  {\tt{github}}\,\cite{gseimgithub}
 \item
  video tutorials for new users
 \item
  course material development based on GSEIM simulation examples
 \item
  additional features such as ``bus" connections, real-time plotting of
  simulation results.
\end{list}

\bibliographystyle{IEEEtran}
\bibliography{ref2}

\end{document}